\documentclass[10pt]{iopart}

\usepackage{iopams}

\usepackage{graphicx}
\usepackage{bm}
\usepackage[urlcolor=blue,colorlinks=true,citecolor=blue,
linkcolor=blue,pdfstartview={FitH},bookmarks=false]{hyperref}
\usepackage{xcolor}

\usepackage{cite}
\usepackage[normalem]{ulem}

\begin{document}

\title[The {\it ab initio} study of unconventional superconductivity \ldots]
{The {\it ab initio} study of unconventional superconductivity in CeCoIn$_{5}$ and FeSe}

\author{Andrzej Ptok}
\address{Institute of Physics, Maria Curie-Sk\l{}odowska University, \\
Plac M. Sk\l{}odowskiej-Curie 1, PL-20031 Lublin, Poland}
\address{Institute of Nuclear Physics, Polish Academy of Sciences, \\
ul. E. Radzikowskiego 152, PL-31342 Krak\'{o}w, Poland}
\ead{aptok@mmj.pl}

\author{Konrad J. Kapcia}
\address{Institute of Physics, Polish Academy of Sciences, \\
Aleja Lotnik\'{o}w 32/46, PL-02668 Warsaw, Poland}
\ead{kapcia@ifpan.edu.pl}

\author{Przemys\l{}aw Piekarz}
\address{Institute of Nuclear Physics, Polish Academy of Sciences, \\
ul. E. Radzikowskiego 152, PL-31342 Krak\'{o}w, Poland}
\ead{piekarz@wolf.ifj.edu.pl}

\author{Andrzej M. Ole\'{s}}
\address{Max Planck Institute for Solid State Research, \\
Heisenbergstrasse 1, D-70569 Stuttgart, Germany}
\address{Marian Smoluchowski Institute of Physics, Jagiellonian University, \\
ul. prof. S. \L{}ojasiewicza 11, PL-30348  Krak\'{o}w, Poland}
\ead{a.m.oles@fkf.mpg.de}

\vspace{10pt}
\begin{indented}
\item[]5 October 2017
\end{indented}

\begin{abstract}
Electronic structure and the shape of the Fermi surface are known
to be of fundamental importance for the superconducting instability
in real materials. We demonstrate that such an instability may be
explored by static Cooper pair
susceptibility renormalized by pairing interaction and present
an efficient method of its evaluation
using Wannier orbitals derived from {\it ab initio} calculation.
As an example, this approach is used to search for an unconventional
superconducting phase of the Fulde--Ferrell--Larkin--Ovchinnikov
(FFLO) type in a heavy-fermion compound CeCoIn$_5$ and an iron-based
superconductor FeSe. The results suggest that the FFLO superconducting
phase occurs at finite magnetic field in both materials.
\end{abstract}

%\pacs{74.20.Pq, 74.70.Xa, 74.78.-w}
\vspace{1pc}

\vspace{1pc}
\submitto{New Journal of Physics}
%\maketitle

\section{Introduction}

In the standard theory of superconductivity developed by Bardeen,
Cooper and Schrieffer (the BCS theory) introduced in 1957~\cite{BCS1,BCS2}
the concept of the Cooper pairs which are in a singlet state with zero
total momentum plays a fundamental role. Shortly after this discovery in
1964 two independent groups proposed an unconventional superconducting
state with the Cooper pairs having non-zero total momentum. First group
of Fulde and Ferrell proposed a state where the Cooper pairs have only
one possible momentum~\cite{FF}, while the second one of Larkin and
Ovchinnikov assumed that pairs with two opposite momenta exist in
superconducting state~\cite{LO}. Thereafter, materials where such
unconventional Fulde--Ferrell--Larkin--Ovchinnikov (FFLO) states could
be realized have been looked for very intensively.

In the absence of an external magnetic field, the Fermi surfaces for
electrons with opposite spins are similar. Then, in conventional BCS-type
superconductor, every electron in state (${\bm k},\uparrow$) is paired
with an electron in state ($-{\bm k},\downarrow$) and the total momentum
of this pair equals zero. Here ${\bm k}$ is the momentum of an electron
with spin $\sigma=\uparrow,\downarrow$.
For this reason one can say that the main source of unconventional
superconductivity of the FFLO-type is a shift of the Fermi surface.
In such a case, an electron in state (${\bm k},\uparrow$) can be paired
with an electron in the state with shifted momentum,
($-{\bm k}$+${\bm q},\downarrow$).
Then, the total momentum of the Cooper pairs in this case is non-zero
and equals ${\bm q}$. Indeed, the main source of the FFLO phase can be
an external magnetic field which leads to the Zeeman effect and to the
occurrence of the shift of the Fermi surfaces for electrons with
opposite spins. However, also other effects such like the mass-imbalance
in the system can lead to the shift of  the Fermi surfaces
\cite{batrouni.huntley.08,batrouni.wolak.09,baarsma.gubbels.10,mathy.parish.11},
which is used in the recent experiments for realisation of the FFLO
phase in ultra-cold fermion gases on optical lattices.

According to the previous theoretical studies, a material where one can
expect a realization of the FFLO phase has to fulfill a few
restrictions and conditions. In such a system the main factor
determining the upper critical magnetic field has to be given by Zeeman
(paramagnetic) effect whereas the orbital (diamagnetic) effect should
be less important or negligible. Such a property can be associated with
the Maki parameter,
$\alpha_{M}=\sqrt{2} H_{c2}^{orb}/H_{c2}^{P}$~\cite{maki.66},
which describes the ratio of the critical magnetic fields at zero
temperature given only by diamagnetic ($H_{c2}^{orb}$) and paramagnetic
($H_{c2}^{P}$) effects. This suggests that the interesting systems in
the context of the occurrence of the FFLO phase are materials where
$\alpha_{M}\geq 1$ (so-called Pauli limited systems).
Moreover, the FFLO phase can be realized at low temperature and high
magnetic field only. Under these conditions, in the absence of orbital
effects, one expects the first order phase transition from the
superconducting FFLO state to the normal (non-superconducting) phase.

In condensed matter relatively large Maki parameter $\alpha_{M}$ is
expected in systems where orbital effects are in natural way negligible,
like in systems with large effective electron mass or in layered
materials. Among the mentioned groups, one may find heavy fermion
systems \cite{matsuda.shimahara.07} (e.g. CeCoIn$_{5}$ which is
described in detail in section~\ref{sec.intro.hfs}) or organic
superconductors \cite{beyer.wosnitza.13} (e.g.
$\beta''$-(ET)$_{2}$SF$_{5}$CH$_{2}$CF$_{2}$SO$_{3}$ \cite{cho.smith.09}, $\lambda$-(BETS)$_{2}$FeCl$_{4}$~\cite{uji.kodama.13}, $\lambda$-(BETS)$_{2}$GaCl$_{4}$~\cite{coniglio.winter.11},
$\kappa$-(BEDT-TTFS)$_{2}$Cu(NCS)$_{2}$
\cite{singleton.symington.00,lortz.wang.07,bergk.demuer.11,
agosta.jin.12,mayaffre.kramer.14,tsuchiya.yamada.15}).
In addition, iron-based superconductors show features suggesting the
possibility of the realisation the FFLO phase (this group of systems is
described in more detail in section~\ref{sec.intro.ibsc}).

\begin{figure}[b!]
\begin{center}
\includegraphics[width=0.75\linewidth]{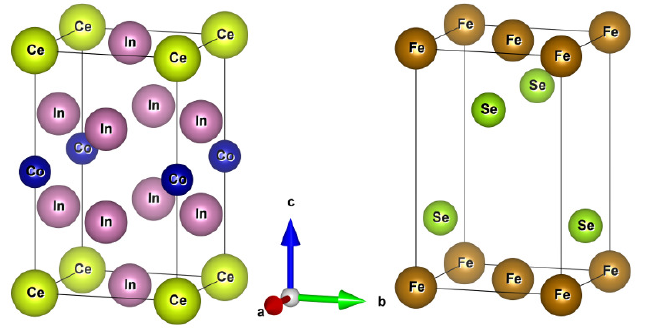}
\end{center}
\caption{
The chemical structure of superconducting heavy-fermion system
CeCoIn$_{5}$ (left) and iron-based superconductor FeSe (right).
The image was rendered using {\sc VESTA} software~\cite{vesta}.
\label{fig.struk}
}
\end{figure}

\subsection{Heavy fermion systems}
\label{sec.intro.hfs}

CeCoIn$_{5}$ is a representative heavy fermion system. This compound
exhibits a layered structure (its crystal structure is presented in the
left panel of figure~\ref{fig.struk}). The heavy fermion systems are
characterized by relatively high Maki parameter, which is estimated as $\alpha_{M}\sim5$~\cite{kumagai.saitoh.06}. The symmetry of
superconducting gap is in agreement with {\it $d$-wave}-type
pairing \cite{aoki.sakakibara.04,izawa.yamaguchu.01}.

A possible existence of the FFLO phase in this compound can be
concluded from several untypical physical properties. Many experiments
provide evidence in favour of the first order phase transition from
the superconducting state to the normal state for the magnetic field
parallel to the main crystallographic directions (${\bm H} \| ab$ and
${\bm H} \| c$). It has been shown in the measurements of a jump of
thermal conductivity~\cite{izawa.yamaguchu.01,capan.bianchi.04},
magnetization~\cite{tayama.harita.02},
penetration depth~\cite{martin.agosta.05},
ultrasound velocity~\cite{watanabe.kasahara.04},
thermal expansion~\cite{oeschler.gegenwart.03},
magnetostriction~\cite{correa.murphy.07},
in NMR experiments~\cite{kumagai.saitoh.06,kakuyanagi.saitoh.05,mitrovic.horvatic.06}
and from the shape of specific heat
\cite{bianchi.movshovich.02,radovan.fortune.03,bianchi.movshovich.03,miclea.nicklas.06}.
Moreover, the experimental results show an existence of incommensurate
spin-density wave (SDW) state in a regime of occurrence of superconducting
state
\cite{kenzelmann.strassle.08,koutroulakis.mitrovic.08,kenzelmann.gerber.10}.
This cannot be explained by a simple assumption of the existence of
superconducting BCS
state~\cite{miyake.08,yanase.08,yanase.sigrist.09,ptok.maska.11}.
The presence of the SDW also enhances a tendency of the system to
stabilize the FFLO phase~\cite{mierzejewski.ptok.09}.

\subsection{Iron based superconductors}
\label{sec.intro.ibsc}

Superconductivity in the iron-based superconductors was discovered in
2008 by Kamichara {\it et al.} in LaFeAsO doped by F at the oxygen
site below 26 K~\cite{kamihara.watanabe.08}. This new class of
high-temperature superconductors has been intensively investigated in
the last few years~\cite{johnston.10,stewart.11,dai.15}.
These materials are characterized by iron-arsenide or iron-selenide layers~\cite{johnston.10,hosono.kuroki.15}. As a consequence, they
have characteristic Fermi surfaces (with coexisting hole and electron
pockets around $\Gamma$ and $M$ point of the first Brillouin zone).
Layered structure and quasi-two-dimensional character of the Fermi
surfaces in many representative systems from the family of iron-based
superconductors make these materials placed in the Pauli limit. 
The Maki parameter in this group of compounds can be estimated as
$\alpha_{M}\sim 1-5$
\cite{yuan.singleton.09,zocco.grube.13,terashima.kihou.13,zhang.singh.15,
fuchs.drechsler.08,fuchs.drechsler.09,khim.lee.11,
cho.kim.11,zhang.jiao.11,kurita.kitagawa.11,chong.williams.14,
jia.cheng.08,tarantini.gurevich.11,kasahara.watashige.14}.
Thus, the first order phase transition from the superconducting to
the normal state and high anisotropy of upper critical magnetic field
have been reported in many experiments for different iron-based
superconductors (e.g. Ba$_{1-x}$K$_{x}$Fe$_{2}$As$_{2}$
\cite{yuan.singleton.09,zocco.grube.13,terashima.kihou.13,zhang.singh.15},
LaO$_{0.9}$F$_{0.1}$FeAs$_{1-\delta}$
\cite{fuchs.drechsler.08,fuchs.drechsler.09},
LiFeAs~\cite{khim.lee.11,cho.kim.11,zhang.jiao.11},
NdFeAsO$_{1-x}$F$_{x}$~\cite{jia.cheng.08},
FeSe$_{1-x}$Te$_{x}$ \cite{tarantini.gurevich.11,kasahara.watashige.14}).

The simplest representative of this group of materials is FeSe (right
panel of figure~\ref{fig.struk}), with the critical temperature $T_c=8$~K
\cite{hsu.luo.08}. It undergoes a structural transition from the
tetragonal to the orthorhombic phase at $T_{s}\simeq 87$ K with no
magnetic order~\cite{mcqueen.huang.09,mcqueen.williams.09}, but with
emerging nematic electronic structure~\cite{watson.kim.15}. On the other
hand, the calculations based on density functional theory (DFT) predict
magnetic states both for FeSe~\cite{li.zhu.09} and FeSe$_{1-x}$ bulk
materials~\cite{kumar.kumar.10}. They indicate the proximity of FeSe to
the magnetic instability, which is not observed due to electron
correlations beyond the DFT~\cite{backes.jeschke.15,hirayama.misawa.15}.
Possible consequences of this proximity have
been debated recently and exotic states with the hidden magnetic
order have been proposed: antiferroquadrupolar spin order \cite{yu.si.15}
and nematic quantum paramagnetic phase \cite{wang.kivelson.15}.
The absence of magnetic order in FeSe makes this material a good
candidate to test the possibility of electron-pairing mechanism of
superconductivity, which is still unknown in iron-based superconductors
\cite{hirschfeld.horshunov.11,wang.lee.11,hosono.kuroki.15}.
In fact, it is likely that magnetic interactions at orbital
degeneracy play an important role \cite{Nic11}.
The coexistence of superconducting and antiferromagnetic orders in
some of iron-based superconductors suggests the existence of
superconductivity with $s_{\pm}$ symmetry~\cite{mazin.singh.08} where
the superconducting gap has opposite signs on the Fermi surface around
the $\Gamma$ and $M$ points of the first Brillouin zone.
However, for nonmagnetic FeSe materials this hypothesis is still under debate~\cite{fan.zhang.15,urata.tanabe.16}.

In the context of the present paper, there are very important
observations of the phase transition in high magnetic field
(above $13.5$ T) and at low temperatures (below $1$ K) in pure single
crystals of superconducting FeSe reported by Kasahara {\it et al.}
\cite{kasahara.watashige.14}. Namely, in the strongly spin-imbalanced
state the zero-resistivity (superconducting) state has been observed
together with an additional phase transition, what can be interpreted
as the prediction of an occurrence of the FFLO phase.

\subsection{Motivation}

As we showed above, relatively large number of various chemical
compounds manifest properties which are typical for systems where one
can expect the FFLO phase occurrence. Their studies enforce a quest for
new theoretical techniques. In this paper, we propose a method for
studying the properties of superconducting states using a combination of
the {\it ab initio} (DFT) and the Cooper pair susceptibility calculations.
The use of DFT leads us to design realistic models to capture the
relevant part of the electronic structure of materials, while the Cooper
pair susceptibility calculations show a tendency of the systems towards
different types of superconductivity without defining its source.
The theoretical background is described in detail in section
\ref{sec.theory}, whereas the methods used are explained in
\ref{sec.method}. Numerical results and their discussion are presented
in section~\ref{sec.num.dis} for two exemplary systems:
a heavy fermion compound CeCoIn$_{5}$ and
an iron-based superconductor FeSe.
The summary and general
discussion are presented in section~\ref{sec.sum}.

\section{Theoretical background}
\label{sec.theory}

In a general case the band structure of the system can be represented
by the non-interacting Hamiltonian in the diagonal form:
\begin{eqnarray}
\label{eq.hamnoint} \mathcal{H}_{0}
= \sum_{\varepsilon{\bm k}\sigma} E_{\varepsilon{\bm k}\sigma}^{}
c_{\varepsilon{\bm k}\sigma}^{\dagger} c_{\varepsilon{\bm k}\sigma}^{},
\end{eqnarray}
where $c_{\varepsilon{\bm k}\sigma}^{\dagger}$
($c_{\varepsilon{\bm k}\sigma}^{}$) is creation (annihilation) operator
(in momentum representation) of an electron in band $\varepsilon$ with
momentum ${\bm k}$ and spin $\sigma$. Similarly,
$E_{\varepsilon{\bm k}\sigma}$ denotes the band energy for an
electron from band $\varepsilon$ with momentum ${\bm k}$ and spin
$\sigma$. For nonmagnetic states considered below these energies are
equal for both directions of electron spin $\sigma=\uparrow,\downarrow$.

To study an unconventional superconducting state, we define the static
Cooper pair (superconducting) susceptibility
$\chi_{\varepsilon}^{\Delta}( {\bm q})$
\cite{mierzejewski.ptok.09,ptok.crivelli.13,januszewski.ptok.15,piazza.zwerger.16}:
\begin{eqnarray}
\chi_{\varepsilon}^{\Delta} ( {\bm q} ) \equiv
\lim_{\omega \rightarrow 0 } \frac{-1}{N} \sum_{ij}
\exp \{ i {\bm q} \cdot ( {\bm i} - {\bm j} ) \}
\langle \langle \hat{\Delta}_{\varepsilon} ( {\bm i} ) |
\hat{\Delta}_{\varepsilon}^{\dagger} ( {\bm j} ) \rangle \rangle^{r}_{\omega} ,
\label{eq.chi}
\end{eqnarray}
where $N$ is the number of sites,
$\langle \langle \cdots \rangle \rangle^{r}_{\omega}$ is the
retarded Green's function and
$\hat{\Delta}_{\varepsilon} ( {\bm i} ) = \sum_{\bm j}
\vartheta ( {\bm j} - {\bm i} ) c_{\varepsilon{\bm i}\uparrow}
c_{\varepsilon{\bm j}\downarrow}$ defines the intraband Cooper pair
annihilation operator in the real space at site ${\bm i}$,
while the vector ${\bm q}$ is total momentum of this pair.
The factor $\vartheta ( {\bm j}-{\bm i})$ defines the type of the
pairing interaction in the real space and corresponds to the
symmetry of the order parameter in the momentum space
\cite{ptok.crivelli.13}. For example, for the on-site $s$-wave pairing
it is given as $\vartheta ( {\bm j} - {\bm i} ) = \delta_{ij}$,
whereas for the $d$-wave pairing
$\vartheta({\bm j}-{\bm i})=\delta_{i\pm\hat{x},j}-\delta_{i\pm\hat{y},j}$,
where $\hat{x}$ and $\hat{y}$ are unit vectors in the $x$- and
$y$-direction of the lattice.

\begin{figure}[b!]
\begin{center}
\includegraphics[width=0.5\linewidth]{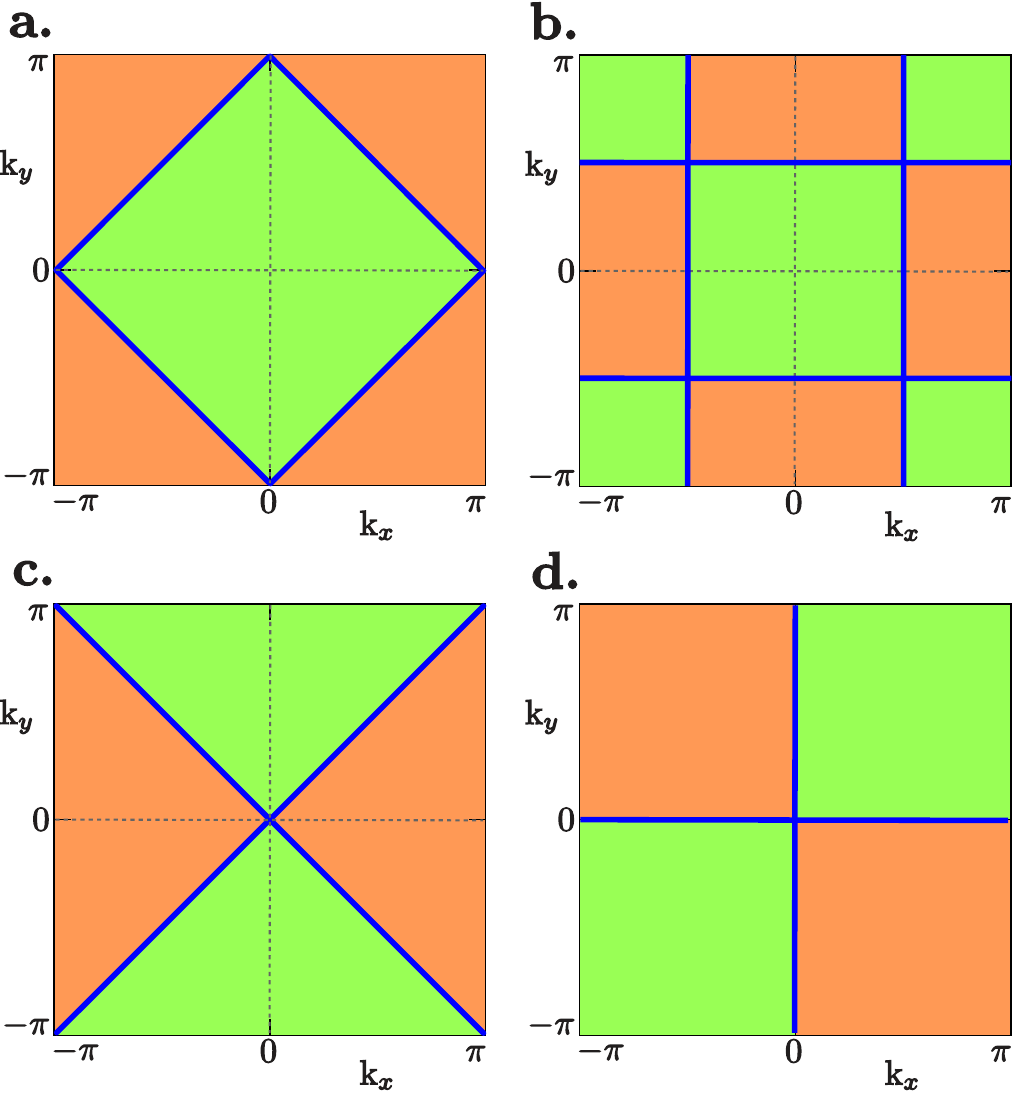}
\end{center}
\caption{
The signs and nodal lines of the superconducting gap for different
symmetries:
(a) $s_{x^{2}+y^{2}}$,
(b) $s_{x^{2}y^{2}}$ ($s_{\pm}$),
(c) $d_{x^{2}-y^{2}}$, and
(d) $d_{x^{2}y^{2}}$.
The background colour corresponds to the sign of the energy gap
(positive and negative shown by orange and green, respectively).
Solid blue lines are nodal lines.
\label{fig.sym}
}
\end{figure}

In the momentum space, the static Cooper pairs susceptibility can be
rewritten in a form
\begin{eqnarray}
\label{eg.chi.momentum}
\chi_{\varepsilon}^{\Delta} ( {\bm q} ) =
\lim_{\omega \rightarrow 0 } \frac{-1}{N} \sum_{{\bm k} {\bm l}}
\eta ( - {\bm k} + {\bm q} ) \eta ( \bm l )
{\bm G}_{\varepsilon} ( {\bm k} , {\bm l} , {\bm q} , \omega ) ,
\end{eqnarray}
where the Green's function is
\begin{eqnarray}
\nonumber {\bm G}_{\varepsilon} ( {\bm k} , {\bm l} , {\bm q} , \omega )
\!&\equiv& \langle \langle  c_{\varepsilon{\bm k}\uparrow}^{}
c_{\varepsilon,-{\bm k}+{\bm q}\downarrow}^{} |
c_{\varepsilon,-{\bm l}+{\bm q}\downarrow}^{\dagger}
c_{\varepsilon{\bm l}\uparrow}^{\dagger} \rangle \rangle_{\omega} = \\
&=& \delta_{{\bm k}{\bm l}}\,\frac{ f( -E_{\varepsilon{\bm k}\uparrow} )
- f( E_{\varepsilon,-{\bm k}+{\bm q}\downarrow} ) }
{ \omega-E_{\varepsilon{\bm k}\uparrow}
-E_{\varepsilon,-{\bm k}+{\bm q}\downarrow} },
\end{eqnarray}
$f(\omega)=1/\{ 1+\exp(\omega/k_{B}T)\}$ is the Fermi-Dirac distribution,
and $\eta ( {\bm k} )$ is the form-factor describing the symmetry of
the order parameter in the momentum space --- it could be equal to $1$,
$2(\cos k_{x}+\cos k_{y})$,    $4\cos k_{x}\cos k_{y}$,
$2(\cos k_{x}-\cos k_{y})$, or $4\sin k_{x}\sin k_{y}$,
for $s$, $s_{x^{2}+y^{2}}$, $s_{x^{2}y^{2}}$ ($s_{\pm}$), $d_{x^{2}-y^{2}}$,
and $d_{x^{2}y^{2}}$ symmetry, as shown in figure~\ref{fig.sym}.
In a case of the cylindrical Fermi surface centered at the $\Gamma$
point and $d$-wave pairing, as consequence, one can expect the nodal
line in the gap (see \ref{fig.sym}(c) and \ref{fig.sym}(d)).
This has measurable consequences as then the properties of the
superconducting phase are quite different
\cite{sigrist.ueda.91,tsuei.kirtley.00}. In the following of this
paper the results presented are obtained for $\eta({\bm k}) = 1$.

From equation~(\ref{eq.chi}) we find that the superconducting
susceptibility, $\chi_{\varepsilon}^{\Delta}({\bm q})$, can be
associated with the effective interaction $\mathcal{H}_{\rm SC}$
describing a superconducting state which in the basis of band operators
$\{c_{\varepsilon{\bm k}\sigma}^{\dagger},
   c_{\varepsilon{\bm k}\sigma}^{}\}$ can be given in a form of
a phenomenological BCS-like term:
\begin{eqnarray}
\mathcal{H}_{\rm SC}= \sum_{\varepsilon{\bm k}} U_{\varepsilon}\left(
\Delta_{\varepsilon{\bm k}}^{\ast} d_{\varepsilon,-{\bm k}
+{\bm q}\downarrow}^{} d_{\varepsilon{\bm k}\uparrow}^{}+ {\rm H.c.}\right),
\end{eqnarray}
where $U_{\varepsilon}\Delta_{\varepsilon{\bm k}}$ denotes an energy-gap
function in a superconducting state (for a given symmetry of the order
parameter in the momentum space)~\cite{ptok.crivelli.13,januszewski.ptok.15}.

The renormalised static Cooper pair susceptibility (taking into account
the effective $U_{\varepsilon}$ interaction) is given in the random
phase approximation as
\begin{equation}
\bar{\chi}_{\varepsilon}^{\Delta}( {\bm q} ) =
\frac{\chi_{\varepsilon}^{\Delta}( {\bm q} )}
{1+U_{\varepsilon} \chi_{\varepsilon}^{\Delta}({\bm q})}\,.
\label{chirpa}
\end{equation}
Here, $\chi_{\varepsilon}^{\Delta}({\bm q})$ is a superconducting
susceptibility (\ref{eq.chi}) at $U_{\varepsilon}=0$ (in the normal state),
while $\bar{\chi}_{\varepsilon}^{\Delta}({\bm q})$ is a susceptibility in
the presence of the effective pairing interaction $U_{\varepsilon}\neq 0$
(in the superconducting state) in a given band $\varepsilon$.

Similarly like in the studies of magnetic properties using the
Lindhard spin susceptibility~\cite{mazin.schmalian.09}, a divergence of
the susceptibility $\bar{\chi}_{\varepsilon}^{\Delta}({\bm q})$
(\ref{chirpa}) suggests the most likely state, while
$U_{\varepsilon}\simeq - 1/\chi_{\varepsilon}^{\Delta}({\bm q})$ can be
treated as a minimal value of interaction $U_{\varepsilon}$ needed to
induce the superconductivity. As shown below, in the absence of the
external magnetic field, $\chi_{\varepsilon}^{\Delta}({\bm q})$ has a
maximal value for ${\bm q}=0$ suggesting a tendency to stabilise the
BCS state. In the context of the FFLO phase, it is of interest to
consider different values of ${\bm q}$ to establish for which one
$\chi_{\varepsilon}^{\Delta}({\bm q})$ has a maximal value in the
presence of the external magnetic field. In such a case, a distinct
maximum for ${\bm q} \neq 0$ can suggest a possibility of realisation
of the FFLO phase in the system. For this reason we calculate and
compare superconducting susceptibilities in the absence as well as in
the presence of the external magnetic field.

Because the physical properties of superconductors are very sensitive
to their electronic structure
(see e.g. \cite{graser.maier.09,kordyuk.12}),
it is crucial to describe it as accurately as possible. In this paper,
we present for the first time the calculations of the superconducting
susceptibility (\ref{eq.chi}) combined with the {\it ab initio} (DFT)
band-structure approach on the examples of two systems:
CeCoIn$_{5}$ and FeSe.

\section{Methods}
\label{sec.method}

\subsection{Conditions imposed on DFT calculations}
\label{sec.dft}

We remark that the previous calculations of the superconducting
susceptibility were performed in the tight binding models
\cite{mierzejewski.ptok.09,ptok.crivelli.13,januszewski.ptok.15,piazza.zwerger.16}.
It was found that a relatively small external magnetic field corresponds
to a {\it large} total momentum of Cooper pairs in the FFLO state.
For example, in a case of the two band model presented by Raghu
{\it et al}~\cite{raghu.qi.08}, the total momentum of the Cooper pairs
was estimated as $\sim 0.3 / a$~\cite{ptok.crivelli.13,ptok.14}, whereas
in a case of three band model proposed by
Daghofer {\it et al}~\cite{daghofer.nicholson.10,daghofer.nicholson.12}
its value was given as $\sim 0.05 / a$~\cite{januszewski.ptok.15}
(where $a$ is lattice constant). On the contrary, for a case of realistic
values of model parameters such like external magnetic field in
iron-based superconductors one can expect a very small total momentum of
Cooper pairs, with value $\sim 0.002 / a$~\cite{ptok.15}. As one can see,
the results for momenta of pairs in the FFLO state depend strongly on
models and parameters used. For this
reason, in the calculation method of the superconducting susceptibility
$\chi_{\varepsilon}^{\Delta}({\bm q})$ used in this work, the DFT
calculation with an extremely dense ${\bf k}$-grid mesh is employed.
The choice of the grid size has a significant influence on the accuracy
of numerical results (what will be shown in the next paragraph).
Therefore, there are significant differences from calculations of spin
\cite{graser.maier.09,ciechan.winiarski.13,ding.lin.13,ciechan.winiarski.15}
or charge~\cite{bosak.chernyskov.14} susceptibilities, where the nesting
vectors are of the same order of magnitude as the Fermi vectors.

For the reasons mentioned above, in this paper the main idea of
calculations of the Cooper pairs susceptibilities is to connect them
with a realistic description of band structures of studied materials
(i.e., CeCoIn$_{5}$ and FeSe), given by equation~(\ref{eq.hamnoint}).
To reproduce the band structure of real materials we have performed DFT
calculations (section~\ref{sec.dft}) on a smaller {\bf k}-grid.
The results are used in order to construct the tight binding model in
Wannier orbital basis (more details can be found in
section~\ref{sec.tbmodel}). The use of the constructed tight binding
model allows us to find the dispersion relation on the sufficiently
dense {\bf k}-grid, which is much denser than those used in DFT
calculations. In the calculations of the Cooper pair susceptibility
presented in this work we will use an effective {\bf k}-grid
(obtained from the tight binding model) with dense $\sim 10^{4}$ times
bigger than that used in typical DFT calculations. This helps us to
increase the accuracy of calculations what is important due to the
following two reasons:
\begin{itemize}
\item[({\it i})]
to calculate the Cooper pairs susceptibility accurately it is necessary
to use relatively dense {\bf k}-grid (while to get the band structure
directly from the DFT calculation on the dense enough {\bf k}-grid
would take a very long time);
\item[({\it ii})]
a relatively small value of the total momentum of the Cooper pairs in
the FFLO phase~\cite{ptok.14,ptok.15} is expected and thus a step in
{\bf k}-grid as small as possible is needed what increases substantially
the number of {\bf k}-points used in calculations.
\end{itemize}
In calculations one has to consider also the external magnetic field
which can be a source of the FFLO phase in studied materials.
For layered materials, where orbital effects can be neglected, it can
be simply done by using the Zeeman term in Hamiltonian (for more
details see section~\ref{sec.zeeman}). In the {\it ab initio}
calculations it is equivalent to non-equal numbers of electrons with
opposite spins.

\subsection{Details of the {\it ab initio} DFT calculations}
\label{sec.dft}

The main DFT calculations have been carried out using the
{\sc Quantum-ESPRESSO} software~\cite{qe,qe2} within the generalized
gradient approximation (GGA)~\cite{gga}. The interactions between the
core and the valence electrons are implemented through the projector
augmented-wave (PAW) method~\cite{paw} employing {\sc PSlibrary}
\cite{pp,pp2} pseudopotentials.
Perdew and Wang (PW91) parameterized GGA functionals \cite{pw91} are
used to describe the exchange-correlation interactions.

\begin{table}[!t]
\centering
\begin{tabular}{c|cc}
\hline
\hline
 & CeCoIn$_{5}$  & FeSe  \\
\hline
space group & P4/mmm & P4/nmm \\
a [\AA] & 4.613 & 3.762 \\
c [\AA] & 7.551 & 4.420 \\
\hline
\hline
\end{tabular}
\caption{\label{tab:param}
Lattice constants for CeCoIn$_{5}$ \cite{moshopoulou.sarrao.02}
and FeSe \cite{hsu.luo.08} used in the present calculations.}
\end{table}

\begin{table}[!b]
\centering
\begin{tabular}{cccc}
\hline
\hline
 CeCoIn$_{5}$ & & FeSe & \\
\hline
Ce & (0,0,0) & & \\
Co & (0,0,$\frac{1}{2}$) & Fe & ($\frac{1}{4}$,$\frac{3}{4}$,0) \\
In1 & ($\frac{1}{2}$,$\frac{1}{2}$,0) & Se & ($\frac{1}{4}$,$\frac{1}{4}$,$z$) \\
In2 & ($\frac{1}{2}$,$\frac{1}{2}$,$z$)&   & \\
\hline
Ref.~\cite{moshopoulou.sarrao.02} & $z=0.3094$ & Ref.~\cite{hsu.luo.08} & $z=0.2402$ \\
this paper & $z=0.3107$ & this paper & $z=0.2436$ \\
\hline
\hline
\end{tabular}
\caption{\label{tab:relax}
Experimental (from \cite{moshopoulou.sarrao.02,hsu.luo.08}) and
relaxed (obtained from DFT calculations presented in this paper)
atomic coordinates for CeCoIn$_{5}$ and FeSe. The upper part of the
table presents atomic coordinates as a function of parameter~$z$
(cf. figure~\ref{fig.struk}).
}
\end{table}

Before we determined the corresponding tight binding models we have
calculated the band structures for both compounds according to the
following scheme:
\begin{itemize}
\item[({\it i})] we optimized the atomic positions of In atoms in
CeCoIn$_{5}$ and Se atoms in FeSe,
\item[({\it ii})] we performed self-consistent calculations of the
charge densities,
\item[({\it iii})] we determined the band structure, i.e., the values
of $E_{{\bm k}\varepsilon\sigma}$ in equation~(\ref{eq.hamnoint}).
\end{itemize}
All DFT calculations of  band structures presented in this paper have
been performed over $10\times 10\times 10$ {\bf k}-points Monkhorst-Pack
mesh \cite{monkhorst.pack.76} and the cut-off energy for the plane waves
expansion was equal to $\sim1088$~eV ($80$~Ry) and $\sim950$~eV ($70$~Ry)
for CeCoIn$_{5}$ and FeSe, respectively.

In the DFT calculation, we have used lattice constants obtained from
the experimental measurements (table~\ref{tab:param}). As a result of
relaxation of In and Se atoms, we have found a very good compatibility
with the experimental atomic coordinates for these atoms
(see table~\ref{tab:relax}).

\subsection{The tight binding model in Wannier orbitals basis}
\label{sec.tbmodel}

\begin{figure}[!b]
\begin{center}
\includegraphics[width=0.75\linewidth]{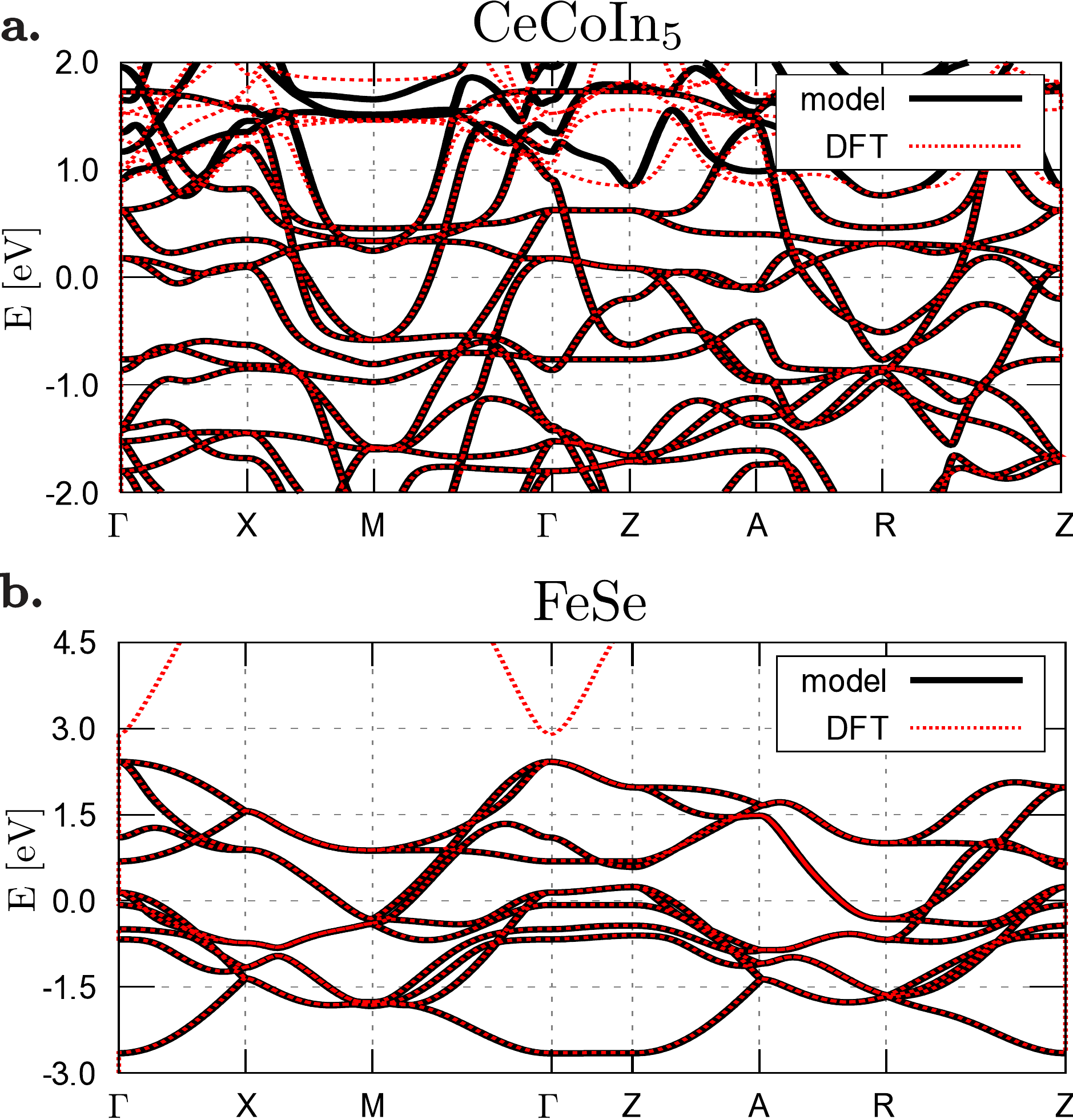}
\end{center}
\caption{
A comparison of electronic band structures found in the {\it ab initio}
DFT calculations (dotted red lines)  and from the tight binding models
in the maximally localized Wannier orbital basis (solid black lines)
along the high symmetry directions in the first Brillouin zone for:
(a) CeCoIn$_{5}$ and
(b) FeSe.
The Fermi level is located at $E=0$~eV.
\label{fig:band}
}
\end{figure}

Using the results of the DFT calculation for electron band structure one
can find the tight binding model in the basis of the Wannier orbitals,
which are located on selected atoms~\cite{marzari.mostofi.12}.
It can be performed by using the {\sc Wannier90} software
\cite{mostofi.yates.08,mostofi.yates.14} which finds the tight binding
model in a base of the maximally localized Wannier functions (MLWF).
As a result of this step one gets a tight binding Hamiltonian of the
electrons with creation operators $\{c_{{\bm R}\mu\sigma}^{\dagger}\}$
on the lattice which is given by
\begin{eqnarray}
\mathcal{H}_{\rm TB}=\sum_{{\bm R},{\bm R}'\sigma}
t_{{\bm R},{\bm R}'}^{\mu\nu}
c_{{\bm R}\mu\sigma}^{\dagger} c_{{\bm R}'\nu\sigma} ,
\end{eqnarray}
where $t_{{\bm R},{\bm R}'}^{\mu\nu}$ are hopping elements between
orbitals $\mu$ and $\nu$ localized on the atoms at sites labeled with
${\bm R}$ and ${\bm R}'$. The matrix of the normal state
(i.e., non-superconducting) Hamiltonian in the momentum space reads:
\begin{eqnarray}
\mathbb{H}_{\mu\nu}({\bm k}) = \sum_{\delta}
\exp \left( {\it i} {\bm k} \cdot \delta \right) t_{\delta}^{\mu\nu},
\end{eqnarray}
where $\delta = {\bm R} - {\bm R}'$ is the distance (in the real space)
of hopping $t_{\delta}^{\mu\nu} \equiv t_{{\bm R},{\bm R}'}^{\mu\nu}$.
A~band structure for given {\bf k}-points, i.e.,
$E_{{\bm k}\varepsilon\sigma}$ appearing in equation~(\ref{eq.hamnoint}),
can be found by diagonalization of the matrix $\mathbb{H}({\bm k})$.

The band structures obtained from the DFT calculations
(cf.~section \ref{sec.dft}) for CeCoIn$_{5}$ and FeSe are shown by
dotted red lines in figures~\ref{fig:band}(a) and \ref{fig:band}(b),
respectively). To describe accurately
superconducting properties of studied compounds it is necessary to have
a good description of states near the Fermi level whereas states far
above (below) are irrelevant. The Wannier-based tight binding model
\cite{marzari.vanderbilt.97,souza.marzari.01,mostofi.yates.14} has been
found from $10\times 10\times 10$ full {\bf k}-point DFT calculation
for random projected $25$ orbitals for CeCoIn$_{5}$ and all $10$
$d$-states in Fe in a case of FeSe.
For both cases the convergence tolerance  of the invariant spread, 
$\sum_{n}\langle r^{2}\rangle_{n} - \langle{\bm r}\rangle^{2}_{n}$, of 
the Wannier function in real space
\cite{marzari.vanderbilt.97,souza.marzari.01,mostofi.yates.14} is equal 
to $10^{-10}$ (in a calculation within the defined energy window).
As the result we find $25$- and $10$-orbital tight binding models for 
the description of CeCoIn$_{5}$ and FeSe, with spread smaller that 
3.1~\AA$^2$ and 5.0~\AA$^2$, respectively.
The band structures obtained from tight binding
models are presented in figure~\ref{fig:band} by black solid lines.

In a general case, in order to find the tight binding model we can
generate the MLWF in an appropriate energy window
\cite{mostofi.yates.08,mostofi.yates.14}. Because for FeSe its band
structure obtained from DFT calculations is given by well separated
bands around the Fermi level, see figure~\ref{fig:band}(b), we can set
the energy window from $-3.0$~eV to $2.7$~eV (with respect to the Fermi
level). Thus the tight binding model found reproduces the DFT data very
well for this iron-based superconductor.
In our approach we restrict ourselves only to the separated states 
(the energy gap is located below and above these states).
As a consequence the bands in this energy range are described by the 
same number of maximally localized orbitals (as number of bands).
These orbitals does not give any contribution to the other band beyond 
the energy window (which does not have any effects on the problems 
studied in the paper, because they are located far away from the Fermi 
level).
A situation is more complicated
for CeCoIn$_{5}$ which has more complex band structure with larger
number of bands, see figure~\ref{fig:band}(a). For a better description
of states near the Fermi level we decided to find a larger number of
the MLWF (containing also fully occupied bands below the Fermi level).
As a consequence, the results reproduced by the tight binding model are
in a good agreement with those obtained within the DFT calculation up
to energy $0.5$~eV above the Fermi level, whereas the unoccupied bands
above it are not perfectly represented. These issues do not affect our
main results which concern superconductivity in the system and thus the
location of states in the neighborhood the Fermi level is the most
significant.

\subsection{External magnetic field}
\label{sec.zeeman}

The main source of the shifted Fermi surfaces can be external magnetic
field given by the Zeeman term. It relatively well describes the
influence of the magnetic field on electrons for a case of Pauli
limited materials where the orbital effect can be neglected. This
situation occurs also for the external magnetic field applied parallel
to the layers of materials, with a weak coupling between them. As a
consequence, the magnetic field has been taken into account in the form
of the Zeeman coupling term added to the Hamiltonian which is given by
\begin{eqnarray}
\label{eq:zeeman}
\mathcal{H}_{\rm mag}=-\frac{1}{2}g\mu_{B}H\sum_{\varepsilon{\bm k}}
(c_{\varepsilon{\bm k}\uparrow}^{\dagger}c_{\varepsilon{\bm k}\uparrow}^{}-
c_{\varepsilon{\bm k}\downarrow}^{\dagger}c_{\varepsilon{\bm k}\downarrow}^{}),
\end{eqnarray}
where $g \simeq 2$ is the gyromagnetic ratio, $\mu_{B}$ is the Bohr
magneton and $H$ is the external magnetic field (in Teslas). Notice that
Zeeman term (\ref{eq:zeeman}) is diagonal and thus it can be easily
included in equation (\ref{eg.chi.momentum}).

Similar approaches associated with addition of terms in the
Hamiltonian have been successfully used for the description of
superconductors, with e.g.
impurities \cite{choubey.berlijn.14,kreisel.choubey.15},
orbital ordering \cite{kraisel.mukherjee.15,mukherjee.kreisel.15}, or
spin-orbit coupling~\cite{kosmider.gonzalez.13,fernandes.vafer.14}.

\section{Numerical results and discussion}
\label{sec.num.dis}

\subsection{Heavy fermion system: CeCoIn$_5$}

\begin{figure}[!t]
\begin{center}
\includegraphics[width=\linewidth]{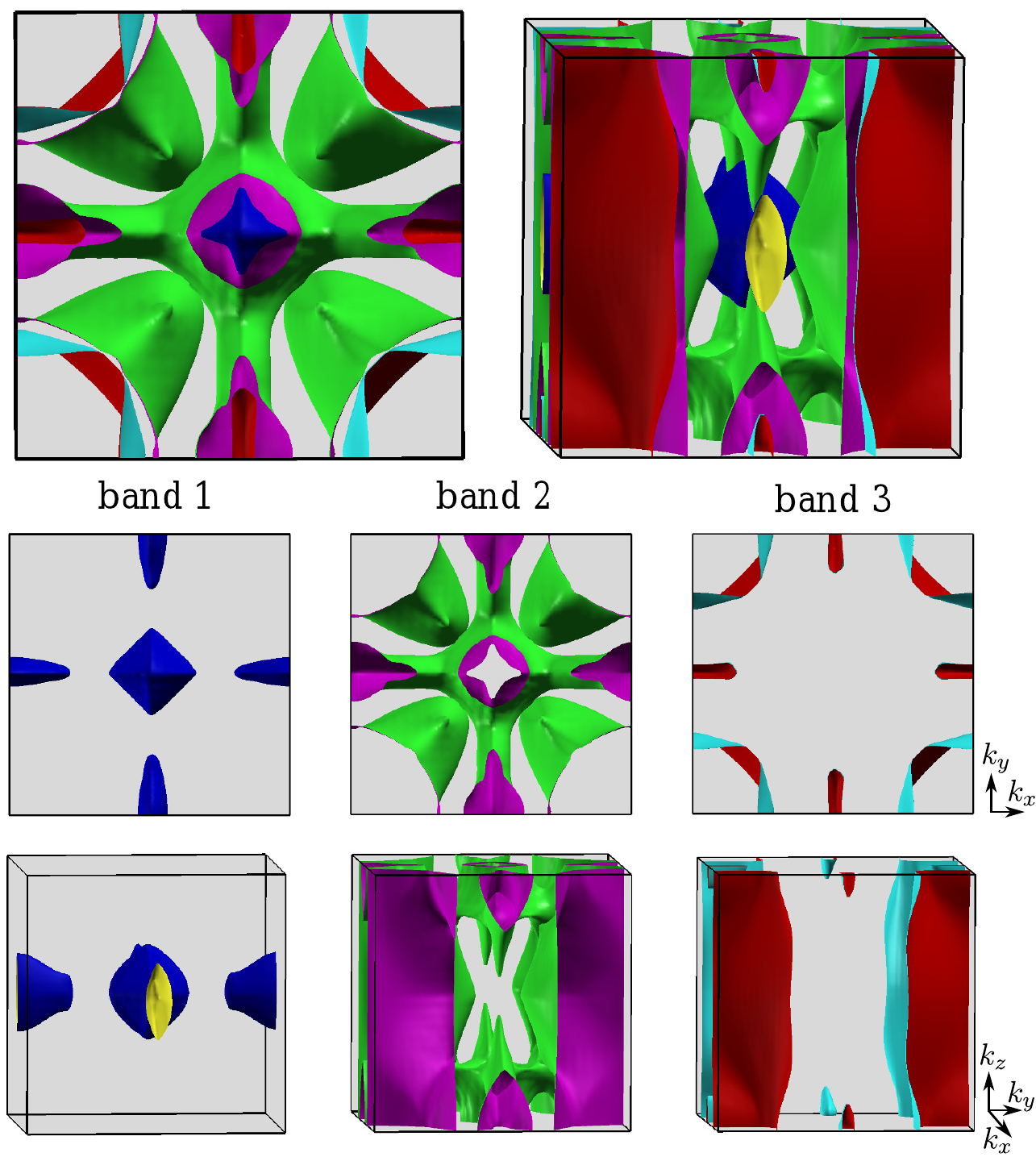}
\end{center}
\caption{
Fermi surfaces for CeCoIn$_{5}$ obtained from the tight binding model.
The first row presents the total Fermi surface originating from all bands
(views from the top --- parallel to the $z$ axis and from the front ---
perpendicularly to the $z$ axis, on left and right panel, respectively).
Panels in the second  and third rows (view from the top and view from
the front, respectively) present the pockets given by each band, separately.
Results  from the tight binding model used in this paper are obtained for
$50 \times 50 \times 50$ {\bf k}-grid mesh. The black boxes denote the
renormalized first Brillouin zone $k_{i}\in (-\pi,\pi)$ ($i=x,y,z$).
\label{fig.fscecoin5}
}
\end{figure}

In this section we discuss the results for the heavy fermion CeCoIn$_5$
compound obtained within the scheme described above. The band structure
obtained for this compound has been presented on the figure~\ref{fig:band}(a).
By using our $25$-orbital tight binding model
we find the Fermi surface of CeCoIn$_5$, which is built from pockets
originating from only three different bands.
It is presented in figure~\ref{fig.fscecoin5}. Our results are in an
agreement with the previous DFT studies presented for this compound \cite{settai.shishido.01,shishido.settai.02,maehira.hotta.03,oppeneer.elgazzar.07,
ronning.zhu.12,polyakov.ignatchik.12}.
The Fermi surface consists of several complex and anisotropic pieces.
There are the edge-centered, more or less cylindrical parts of the
electron-like (two-dimensional) sheets of 2nd and 3rd bands, centered
at $(\pi,\pi,k_{z})$-points and hole-like sheets centered at the
$\Gamma$ point. In this case, the Fermi surface is made by only three
out of $25$ bands. As a consequence, in the following studies of
CeCoIn$_5$ we concern only three of the Cooper pairs susceptibilities
(for every Fermi pocket separately), due to the fact that the  filled
(empty) bands do not affect the superconducting properties of the system.

Next, we are able to calculate the Cooper pairs susceptibility
$\chi_{\varepsilon}({\bm q})$ defined by equation (\ref{eq.chi}) with
or without the magnetic field using a relatively sparse
$50\times 50\times 50$ {\bf k}-grid mesh. As we have stated above, the
location of the maximal value of the Cooper pairs susceptibility in the
momentum space is the most important here because it contains
information about preferred momentum of the Cooper pairs realised in
the system.
In the absence of magnetic field  
maximum values of susceptibilities $\chi_{\varepsilon}({\bm q})$ for
each band $\varepsilon=1,2,3$ are located at the center of the first
Brillouin zone (at the $\Gamma=(0,0,0)$ point), whereas in the presence
of magnetic field they
are located near the $\Gamma$ point and they are not zeros
(at least one $q_\alpha\neq 0$, $\alpha=x,y,x$).

\begin{figure}[t!]
\centering
\includegraphics[width=\linewidth]{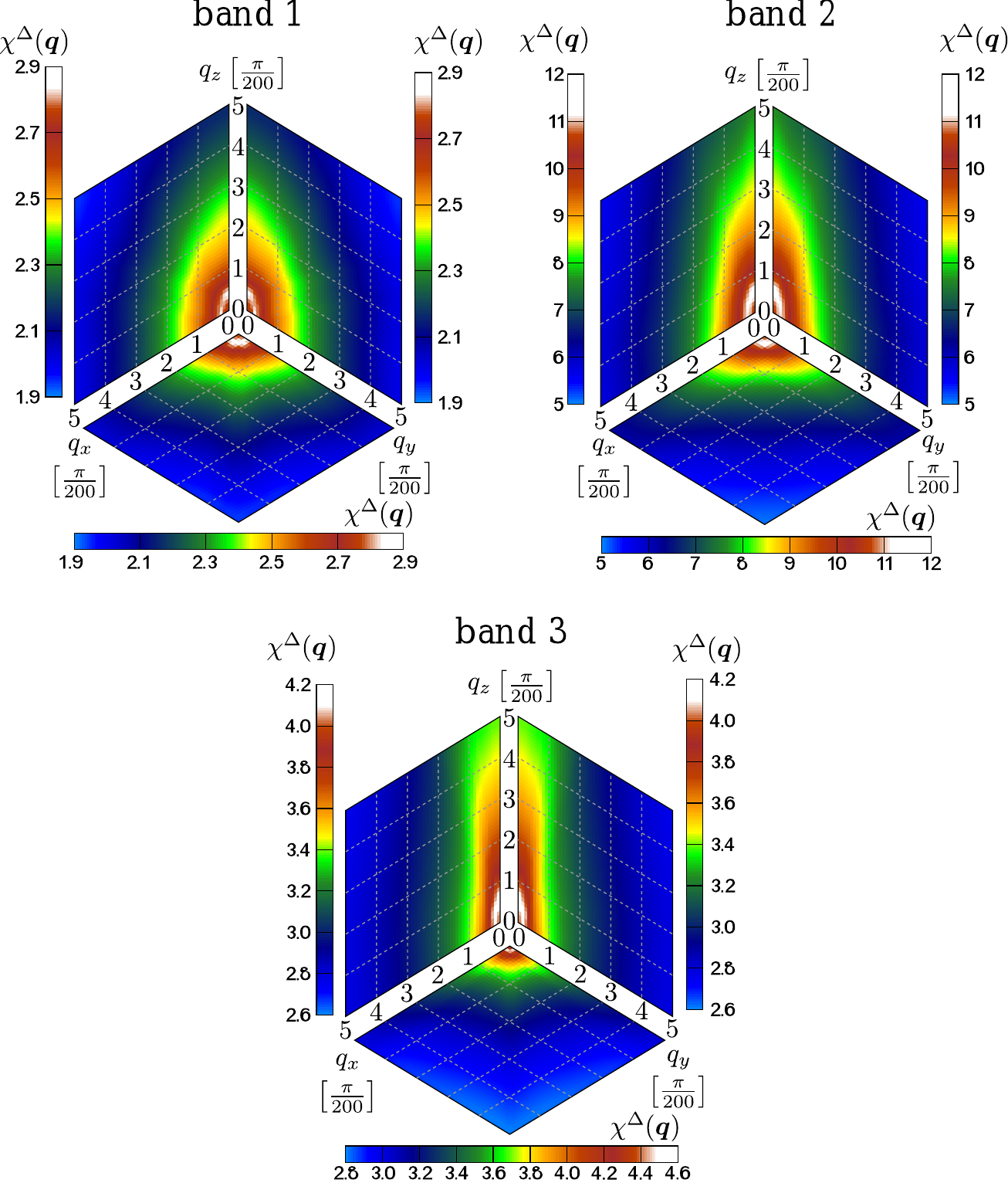}
\caption{
The Cooper pair susceptibility $\chi_{\varepsilon}^{\Delta}({\bm q})$
in the normal state of CeCoIn$_5$ at each band crossing the Fermi level
calculated for momenta within $xy$, $yz$, and $xz$ planes in the
momentum space. The results obtained from tight binding model at
$H=20$~T with $2000\times 2000\times 10$ {\bf k}-grid mesh.
\label{fig.podcecoin5dens}
}
\end{figure}

It should be noted that the result can be visible only for relatively
very large (and nonphysical) magnetic field (here the calculations were
performed for $H=200$~T). In this case, the maxima of
$\chi_{\varepsilon} ({\bm q})$ are located at $|{\bm q}|\sim\pm 2\pi/20$.
It is a consequence of a relatively {\it small} density of the
{\bf k}-grid used in the calculations (i.e., $50\times 50\times 50$
{\bf k}-grid mesh in this example). To obtain more realistic values of
magnetic field at which the FFLO instability can occur, i.e., these from
the FFLO phase regime, $H\approx H_{c}$ which equals approximately
$15$~T, we have to perform calculations using an {\it extremely dense}
{\bf k}-grid.

\begin{figure}[t!]
\begin{center}
\includegraphics[width=\linewidth]{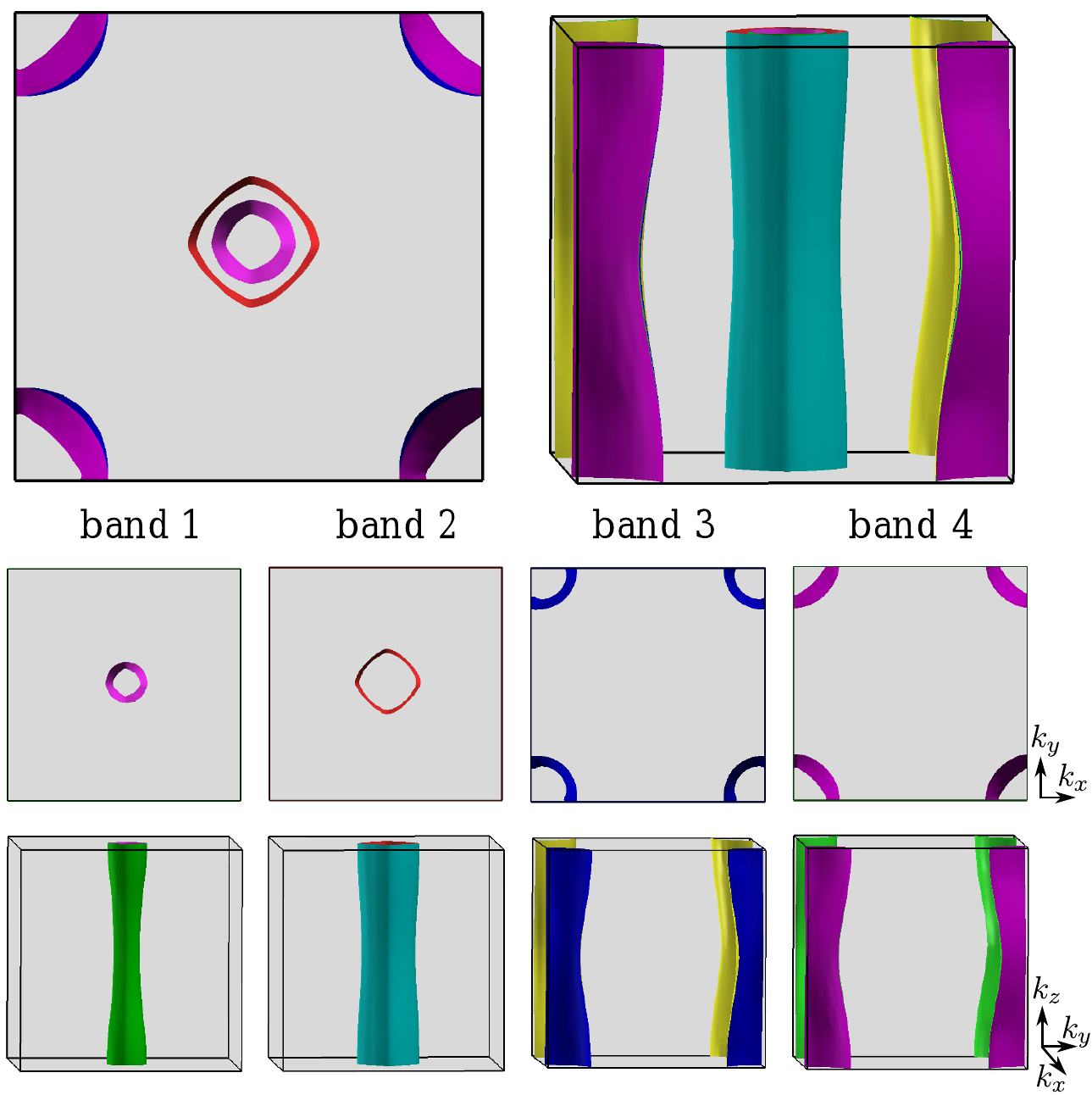}
\end{center}
\caption{
Fermi surfaces for FeSe obtained from the tight binding model.
The first row presents the total Fermi surface (views from the top ---
parallel to $z$-axis and from the front --- perpendicularly to $z$-axis,
on left and right panel, respectively).
Panels in the second and third rows (view from the top and view from
the front, respectively) presents the pockets given by each band,
separately. The results are obtained from the tight binding model used
in this paper for $50 \times 50 \times 50$ {\bf k}-grid mesh.
The black boxes denotes the renormalized first Brillouin zone
$k_{i} \in ( -\pi,\pi )$ ($i=x,y,z$).
\label{fig.fsfese}
}
\end{figure}

Thus, for more accurate description of the system in smaller magnetic
field which may be accessible experimentally, we used the increased
mesh of the {\bf k}-grid in the following way:
\begin{itemize}
\item we find the band structure with a {\it extremely dense} grid
(i.e., $2000 \times 2000$ {\bf k}-point) in a $xy$-plane of the total
momentum of the Cooper pairs {\bf q} using the tight binding
Hamiltonian, and then
\item we {\it sparse the grid} (by 10 ${\bf k}$-points) in the
direction perpendicular to ${\bf q}$.
\end{itemize}
As a result one is able to find locations of maxima of
$\chi_{\varepsilon}^{\Delta}({\bm q})$ with use of
$2000\times 2000\times 10$ {\bf k}-points in the calculations. We
conducted the calculations for the total momenta {\bf q} of Cooper
pairs which are located on $xy$, $xz$ and $yz$ planes of the momentum
space. The results for a case of external magnetic field equal to
$20$~T are shown in figure~\ref{fig.podcecoin5dens}. As one can see
the maximum values of the $\chi_{\varepsilon}^{\Delta}$ are located at
finite values of momenta, at least for the planes considered. Thus one
can conclude that, similarly like previously at $H=200$~T, the maxima
of $\chi_{\varepsilon}^{\Delta}$ occur for non-zero $q$ near the
$\Gamma$ point in each band considered. Let us emphasize that the
maximum of $\chi_{\varepsilon}^{\Delta}$ in the second band has a
relatively large value when it is compared with those of the other bands.

\subsection{Iron based superconductor: FeSe}

In this section we present the results for the bi-atomic FeSe compound
from the family of the iron-based superconductors. The band structure
for the compound have been presented in figure \ref{fig:band}(b). The
four bands compose the Fermi surfaces which is obtained from a
$10$-band tight binding model and it is presented in figure
\ref{fig.fsfese}. Two hole-like bands are centered at the $\Gamma$
point and two electron-like bands at the M point. The calculated band
structure along the $\Gamma$--Z line suggests that, because the
dispersion of the bands is linear in $k_{z}$-direction this system can
be treated as a two dimensional one. This result agrees qualitatively
with the previous calculations~\cite{watson.kim.15,subedi.zhang.08,
singh.12,terashima.kikugawa.14,maletz.zabolotnyy.14,shimojima.suzuki.14,
winiarski.samselczekala.14,leonov.skornyakov.15,lohani.mishra.15}.

\begin{figure}[t!]
\centering
\includegraphics[width=\linewidth]{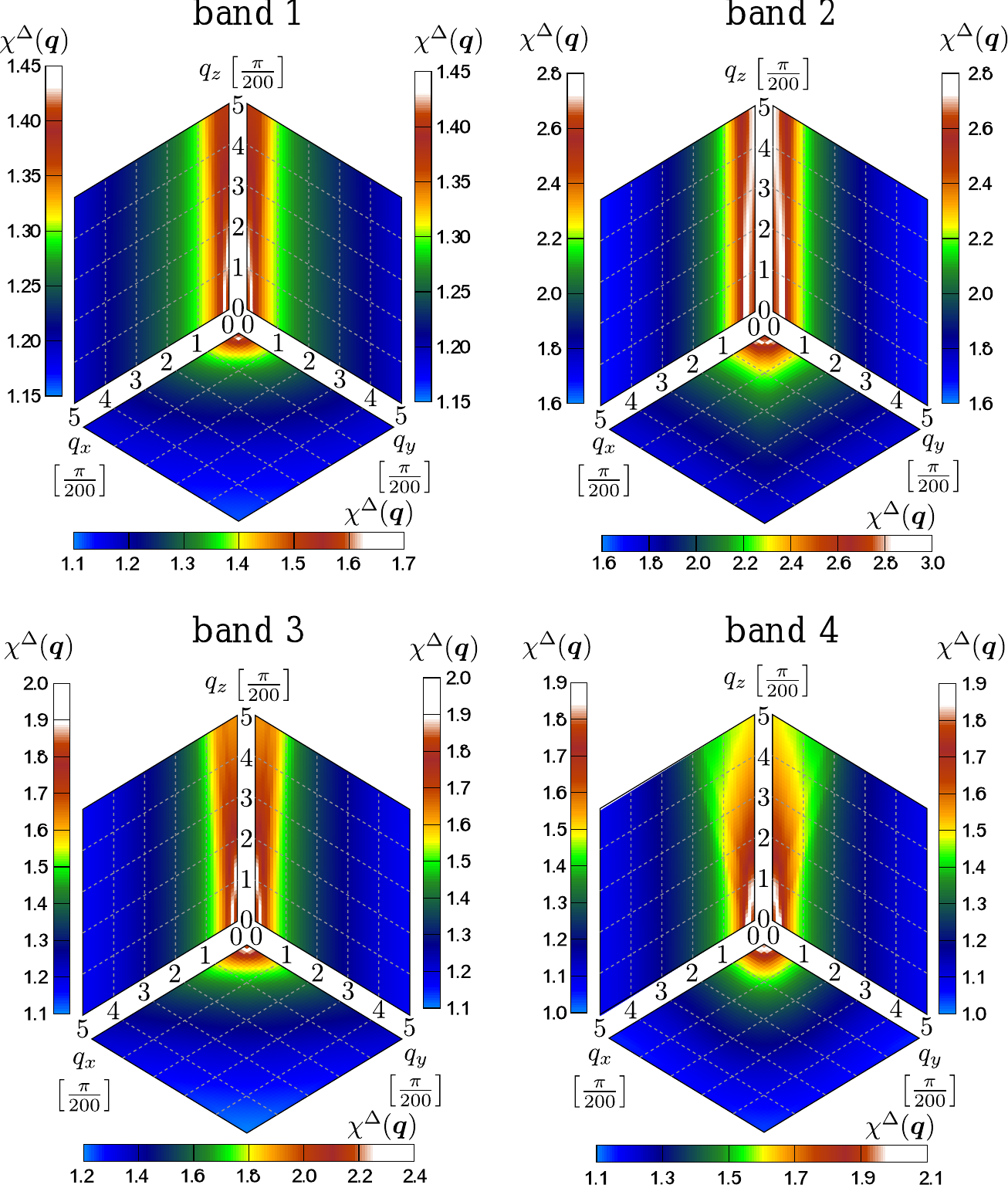}
\caption{
The Cooper pair susceptibility $\chi_{\varepsilon}^{\Delta}({\bm q})$
in the normal state of FeSe at each band crossing the Fermi level
calculated for momenta at $xy$, $yz$, and $xz$ planes of the momentum
space. The results are obtained from the tight binding model at
$H=20$~T with $2000 \times 2000 \times 10$ {\bf k}-grid mesh.
\label{fig.podfesedens}
}
\end{figure}

Similarly like in CeCoIn$_{5}$, in the absence of the external magnetic
field $\chi_{\varepsilon}^{\Delta}({\bm q})$ for FeSe clearly shows a
tendency of the system to realise the homogeneous superconductivity of
the BCS-type in each band. In a case of calculations with the sparse
{\bf k}-grid mesh (i.e., $50\times 50\times 50$) results obtained in a
presence of external magnetic field are similar to those for CeCoIn$_5$.
When one uses too
small accuracy for a grid of {\bf k}-points the results which are
qualitatively different than the BCS (i.e., the results with maxima of
$\chi_{\varepsilon}^{\Delta}({\bm q})$ located not at the $\Gamma$ point),
are possible to find only in extremely large nonphysical magnetic fields.
Thus, similarly like in calculation for the previous heavy-fermion
compound, we use dense grid. In this case in the non-zero magnetic field,
maxima of susceptibilities $\chi_{\varepsilon}^{\Delta}({\bm q})$ in
each band among these building the Fermi surface move to finite values
of the Cooper pair momentum ${\bm q}={\bm q}_{max}\neq0$. The results
for $H=20$~T external magnetic field are shown in figure
\ref{fig.podfesedens} which can be treated as a good approximation of
the real value of the critical magnetic field of this compound.

We would like to emphasize that the DFT predicts low-energy bands which 
deviate quantitatively from the ones observed, e.g. by angle-resolved 
photoemission (ARPES) \cite{watson.yamashita.15,borisenko.evtushinsky.16}.
The both electron and hole Fermi pocket sizes are smaller than those 
obtained from the DFT calculations.
This corresponds to the effective shift of the Fermi level in $\Gamma$ 
and $M$ points~\cite{borisenko.evtushinsky.16}. However, this shift 
should not change qualitatively the results obtained. The correct band 
structure can be found by the dynamical mean field theory (DFT+DMFT) 
calculation \cite{watson.backes.17,yin.haule.14}, but this method gives blurred bands 
(i.e., the bands with finite lifetime) so the results cannot be directly 
used for the calculation of superconducting susceptibility.

\paragraph{Additional remark.}
As it was written in section~\ref{sec.theory},
$1/\chi_{\varepsilon}^{\Delta}({\bm q})$ can be connected with a
minimum value of interaction $U_{\varepsilon}$ which is needed to
induce superconductivity with the Cooper pairs with momentum ${\bm q}$
in the system. In a case of the BCS state where ${\bm q} = 0$, a value
of $\chi_{\varepsilon}^{\Delta}(0)$ can be found using a sparse
{\bf k}-grid mesh. A situation is more complicated when one considers
the FFLO state, where ${\bm q}\neq 0$, but $|{\bm q}|$ is relatively
small. Such conditions can be expected for experimentally available
({\it realistic}) magnetic fields. High critical magnetic fields of the
order of $100-200$~T have been found in many compounds, even for some
from the group of iron-based superconductors~\cite{gurevich.11}. But
such large upper critical fields are given rather by the orbital effect,
what excludes the existence of the FFLO phase in such systems. The
second issue which should be stressed is the fact that a maximum value
of $\chi_{\varepsilon}^{\Delta}$ in the presence of magnetic field is
smaller than that obtained in the absence of the field
(they occur for different momenta $\bm q$: $\bm q=0$ and $\bm q\neq 0$,
respectively).
As a consequence, the magnitude of the effective pairing interaction
leading to the BCS phase,
$U_{\varepsilon}\sim -1/\chi_{\varepsilon}^{\Delta}({\bm q}=0)$,
is bigger than the effective pairing interaction which triggers the FFLO
state (because
$\chi_{\varepsilon}^{\Delta}({\bm q}=0)<
\chi_{\varepsilon}^{\Delta}({\bm q}\neq 0)$).
Thus, for $U_{\varepsilon}$ the FFLO phase can be induced whereas the
BCS-type superconductivity cannot be realised.

\section{Summary}
\label{sec.sum}

In this paper we proposed a method for studying the properties of
superconducting states using a combination of the {\it ab initio}
density functional theory (DFT)
and the Cooper pair susceptibility calculation. We have applied this
method to study a tendency to stabilise the unconventional
superconductivity of the Fulde--Ferrell--Larkin--Ovchinnikov (FFLO)
type in two compounds: CeCoIn$_5$ and FeSe. Using the realistic band
structure determined by the DFT calculations we have derived the
respective tight binding models and obtained the static Cooper pair
susceptibility. We show that the conventional BCS-type superconductivity
is favoured for each band in the absence of magnetic field whereas the
presence of magnetic field can stabilize the
inhomogeneous FFLO phase. Similarly like in the previous theoretical
works \cite{ptok.15}, the total momenta of Cooper pairs are very small.

The present calculations show that
in a general case the total momentum of Cooper pairs ${\bm q}$
changes nonlinearly with magnetic field~\cite{ptok.15}. Moreover, we
can expect that the FFLO superconducting phase can survive for a case
of the band with strong linearity near the Fermi energy because hot-spot
points ${\bm k}$ and $(-{\bm k}+{\bm q})$ can be found there what causes
a growth of $\chi_{\varepsilon}^{\Delta}({\bm q})$ for the non-zero
total momentum of Cooper pairs (${\bm q} \neq 0$). The total momenta of
the Cooper pairs which can be realised in this system are in good
agreement with the previous studies~\cite{ptok.15}.

The calculations have also been performed for other symmetries of the
superconducting gap displayed in figure \ref{fig.sym} as well, i.e.,
$s_{x^2+y^2}$, $s_{x^2y^2}$ ($s_{\pm}$), $d_{x^2-y^2}$, and
$d_{x^2y^2}$. The results are qualitatively similar to these discussed
above for the $s$-wave superconductivity and give enhanced Cooper pair
susceptibility which favours the FFLO phase.
In the case of the $d$-type symmetries, the numerical
results of $\chi_{\varepsilon}^{\Delta}({\bm q})$ found in the presence
of external magnetic field are not distinct and maxima of
$\chi_{\varepsilon}^{\Delta}({\bm q})$ are fuzzy. Moreover, only ground
state calculations for various
superconducting states can resolve the question concerning
the realisation of a particular phase~\cite{ptok.crivelli.13}.

In summary,
we would like to emphasize that the pair susceptibility calculations
presented here can be a useful tool to investigate a tendency of the
system to stabilize some kind of superconducting phase
(in this work we discuss the BCS and FFLO phases).
The calculations for different symmetries of the superconducting gap
were performed and they gave results which are qualitatively similar to
those obtained for the case of the $s$-wave symmetry. Indeed, they give
very similar dependence of the superconducting susceptibility on the
Cooper pair momenta and the external magnetic field. Therefore the
results obtained for the $s$-wave symmetry are generally valid and we
conclude that the electronic structures of both CeCoIn$_5$ and FeSe
support the existence of the FFLO superconducting phase.

\ack

We thank Mario Cuoco, Alexander N. Yaresko and Pawe\l{} T. Jochym
for insightful comments.
This work was supported by Large Infrastructures
for Research, Experimental Development and Innovations project
"IT4 Innovations National Supercomputing Center --- LM2015070"
of the Czech Republic Ministry of Education, Youth and Sports.
We kindly acknowledge support by Narodowe Centrum Nauki
(NCN, National Science Center, Poland),
Projects No.~2016/20/S/ST3/00274 (AP) and
2012/04/A/ST3/00331 (AMO and PP).

\section*{References}

\bibliography{biblio}

\end{document}